\documentclass[english,aps, twocolumn]{revtex4}
\usepackage[T1]{fontenc}
\usepackage[latin9]{inputenc}
\usepackage{amstext}
\usepackage{graphicx}
\usepackage{amssymb}

\makeatletter
\@ifundefined{textcolor}{}
{%
 \definecolor{BLACK}{gray}{0}
 \definecolor{WHITE}{gray}{1}
 \definecolor{RED}{rgb}{1,0,0}
 \definecolor{GREEN}{rgb}{0,1,0}
 \definecolor{BLUE}{rgb}{0,0,1}
 \definecolor{CYAN}{cmyk}{1,0,0,0}
 \definecolor{MAGENTA}{cmyk}{0,1,0,0}
 \definecolor{YELLOW}{cmyk}{0,0,1,0}
 }

\makeatother

\usepackage{babel}

\begin{document}

\title{Localized States and Resultant Band Bending in Graphene Antidot Superlattices}

\author{Milan Begliarbekov$^{1}$, Onejae Sul$^{2}$, John J. Santanello$^{1}$,
Nan Ai$^{1}$, Xi Zhang$^{1}$, Eui-Hyeok Yang$^{2}$, and Stefan
Strauf$^{1}$}

\affiliation{1 Department of Physics and Engineering Physics, Stevens Institute
of Technology, Hoboken NJ, USA}

\affiliation{2 Department of Mechanical Engineering, Stevens Institute of Technology,
Hoboken NJ, USA}
\begin{abstract}
We fabricated dye sensitized graphene antidot superlattices with the
purpose of elucidating the role of the localized edge state density.
The fluorescence from deposited dye molecules was found to strongly
quench as a function of increasing antidot filling fraction, whereas
it was enhanced in unpatterned but electrically back-gated samples.
This contrasting behavior is strongly indicative of a built-in lateral
electric field that accounts for fluorescence quenching as well as
p-type doping. These findings are of great interest for light-harvesting
applications that require field separation of electron-hole pairs.
\end{abstract}


\maketitle
Graphene, a two dimensional monolayer of carbon atoms arranged in
a hexagonal lattice has been recently isolated \cite{Novoselov2004}
and shown to exhibit excellent electrical \cite{Lin2010,Bolotin2008},
thermal \cite{ABalandin2008}, mechanical \cite{Hone_Durability}
and optical \cite{Opto_Review} properties. Electron transport has
been studied extensively in single and few-layer graphene sheets \cite{Novoselov2007,Neto2009},
while optoelectronic properties and light matter interaction in nanostructured
graphene gain increasingly more interest in the research community,
in particular since the advent of first ultrafast graphene photodetectors
\cite{Graphene_Photo}. Single layer graphene absorbs only 2.3\% of
the incident radiation in the visible spectrum \cite{Optical_Transparency},
consequently, efficient photocarrier separation within graphene becomes
particularly important. In order to create a built-in electrical field
that facilitates carrier separation silicon based technology relies
on the pn-junction that is created by doping the silicon lattice.
Physical doping of graphene has been previously achieved by addition
of extrinsic atomic \cite{Atomic_Doping,N_Doping_2} or molecular
\cite{Molecular_Doping,Molecular_Doping_2} species either by adsorption
or intercalation into the graphene lattice \cite{N_Doping_2,Nitrogen_Doping}.
A potentially simpler way to make graphene a viable material for optoelectronics
can be achieved by utilizing lateral electric fields created by Schottky
barriers near the source and drain metal contacts \cite{Contacts_1,Graphene_Photo,Graphene_Photo_2},
as was previously done in carbon nanotubes \cite{Avouris_CNT_Review}.
In the presence of such metal contacts it was also observed that nanotube
fluorescence can be significantly enhanced \cite{Hong_CNT}. While
graphene does not display any exciton emission, quantum dots placed
on unpatterned graphene were recently shown to undergo strong fluorescence
quenching, which is indicative of energy transfer from the quantum
dot exciton oscillator into graphene \cite{Energy_Transfer}. Such
hybrids between graphene and light harvesting molecules can potentially
overcome the low absorption efficiency of bare graphene.

Nanostructured graphene offers further possibilities to explore light
harvesting and carrier separation. Of particular interest are the
so called antidot superlattices, i.e., lattices comprized of a periodic
arrangement of perforations in the underlying graphene structure.
These superlattices were predicted to posses a nonnegligible magnetic
moment \cite{Wimmer_antidot}, a small band gap \cite{Bai_nanomesh,Liang_BandGap,Kim_Nanoperf,Sinitskii_Patterning}
that can be controlled by the antidot filling fraction \cite{antidot_polaron,Antidot_quasiparticles},
and Peierls type electron-hole coupling that leads to polaronic behavior
\cite{antidot_polaron}. In a previous work, Heydrich et al., showed
that the introduction of an antidot superlattice results in the stiffening
of the G-Band in Graphene's Raman spectrum, as well as an energetic
shift of the G and G'-Bands commensurate with p-type doping \cite{Heydrich_antidot_apl}.
Furthermore, recent theoretical predictions show that the periphery
of graphene possesses a nonnegligible density of states $N_{edge}$
that is spatially localized at the edges and is distinct from the
bulk states $N_{bulk}$ that are present in graphene's interior regions.
Consequently, antidot superlattices provide a natural framework for
studying these states and their properties, since the edge states
in these systems coexist with the bulk states, unlike in dot lattices,
where the ratio of edge to bulk states is small.

\begin{figure}
\includegraphics[scale=0.45]{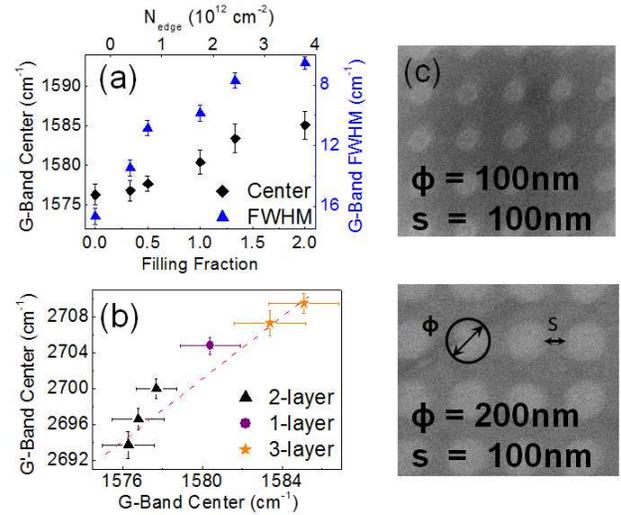}

\caption{(a) The energetic shift (black diamonds) and broadening (blue triangles)
of graphene's G-band as a function of the antidot filling fraction.
(b) Positive correlation of the energetic shifts of the G' and G bands
on different mono, bi, and tri layer samples, showing effective p-doping.
(c) Example SEM images of antidot lattices with different filling
fraction.}

\end{figure}

Here we report an electro-optical study of dye sensitized graphene
antidot superlattices with the purpose of elucidating the role of
the localized edge state density on its light-harvesting properties.
The amount of p-type doping introduced by the edge states is quantified
for various antidot filling fractions using confocal $\mu$-Raman
spectroscopy and transport measurements. We show that the fluorescence
from deposited dye molecules strongly quenches in linear proportion
to the antidot filling fraction, whereas it was enhanced in the presence
of free carriers in unpatterned but electrically back-gated samples.
This contrasting behavior is strongly indicative of a built-in lateral
electric field that accounts for fluorescence quenching as well as
p-type doping and the observed Raman signatures. Our study provides
new insights into the interplay of localized edge states in antidot
lattices and the resulting band bending, which are critical properties
to enable novel applications of nanostructured graphene for light
harvesting and photovoltaic devices.

\section{Results and Discussion}

\subsection{Antidot Superlattices}

Graphene flakes used in these experiments were prepared by micromechanical
exfoliation of natural graphite onto a degenerately doped p$^{+\textrm{ }+}$
Si wafer with a thermally grown 90 nm SiO$_{2}$ dielectric. Layer
metrology was subsequently performed using confocal $\mu$-Raman spectrometry
in order to identify mono, bi, and tri-layer graphene flakes \cite{Begliarbekov2010,Graphene_Raman}.
Following the initial characterization, various antidot superlattices
were etched onto the flakes using electron beam lithography. Figure
1c shows two exemplary lattices with different filling fractions $F=\phi/s$
of antidots, where $\phi$ is the antidot diameter, and $s$ is their
separation. In accordance with previous experimental results \cite{Raman_E_Field_Tuning,Heydrich_antidot_apl,BO_Breakdown},
the corresponding Raman spectra display an energetic shift and linewidth
narrowing of the G-band with increasing filling fraction, as shown
in Fig 1a. The G band, which occurs at \textasciitilde{}1580 cm$^{-1}$
arises from doubly degenerate iTO and iLO phonon modes which possess
$E_{2g}$ symmetry. The observed stiffening (from 16.7 cm$^{-1}$
to 6.6 cm$^{-1}$) can be understood in terms of the Landau damping
of the phonon mode, while the energetic shift arises from a renormalization
of the phonon energy \cite{Raman_E_Field_Tuning,Ferrari2007}. Furthermore,
the energetic shift of the G-band is positively correlated with the
shift of the G'-band, as shown in Fig. 1b, which is indicative of
an effective p-doping of the underlying graphene layer\cite{Gated_Raman,Raman_Doping_Stampfer}.
In contrast, a negative correlation in the energetic shifts of the
G and G' bands would imply n-doping. 

In order to correlate shift and stiffening of the G-band in antidot
superlattices to an underlying carrier density, we fabricated electrically
contacted devices without an antidot lattice, as shown schematically
in Fig. 2b. Using the electrical field effect of the back gate, the
sheet carrier density $\Delta n_{s}$ was modulated and the stiffening
and energetic shift of the G-band in the unpatterned samples was used
to estimate the edge state density in the antidot superlattice (see
supporting online material). From these data the amount of p-doping
in the antidot samples was determined to reach up to $4\times10^{12}$
cm$^{-2}$ at a filling fraction of two (top axis in Fig.1a), and
was not found to depend on the number of graphene layers as shown
in Fig. 1b. The large amount of effective p-doping is rather remarkable
since neither extrinsic dopants, nor an external gate potential were
applied to the antidot samples. 

\begin{figure}
\includegraphics[scale=0.42]{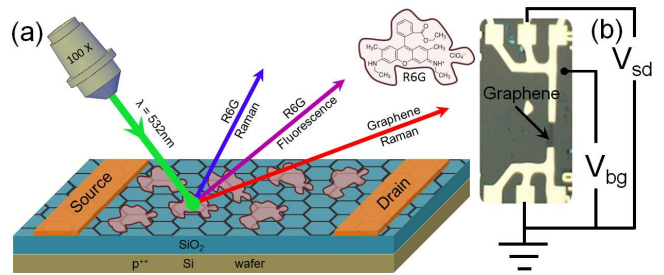}

\caption{(a) Schematic of the spatially resolved confocal $\mu$-Raman experiment,
showing the excitation beam ($\lambda_{0}=532$ nm) and the three
optical signals, R6G Raman, R6G fluorescence and graphene Raman, that
were monitored during these experiments in both, electrically gated
and antidot devices. (b) An example of an electrically contacted graphene
device used in these experiments. }

\end{figure}

Furthermore, in order to investigate the microscopic origin of the
observed p-doping we fabricated graphene-dye hybrids. Both, antidot
flakes and electrically contacted devices were soaked in a 15 nanomol
solution of Rhodamine 6G (R6G), as shown schematically in Fig. 2a.
In these experiments, the R6G Raman peaks, the R6G fluorescence, and
the Raman signal from graphene were monitored as a function of the
antidot filling fraction $F$ as well as different backgate and source-drain
biases on the unpatterned flakes. In the subsequent discussion, we
first focus on the R6G fluorescence signal.

\begin{figure*}
\includegraphics[scale=0.75]{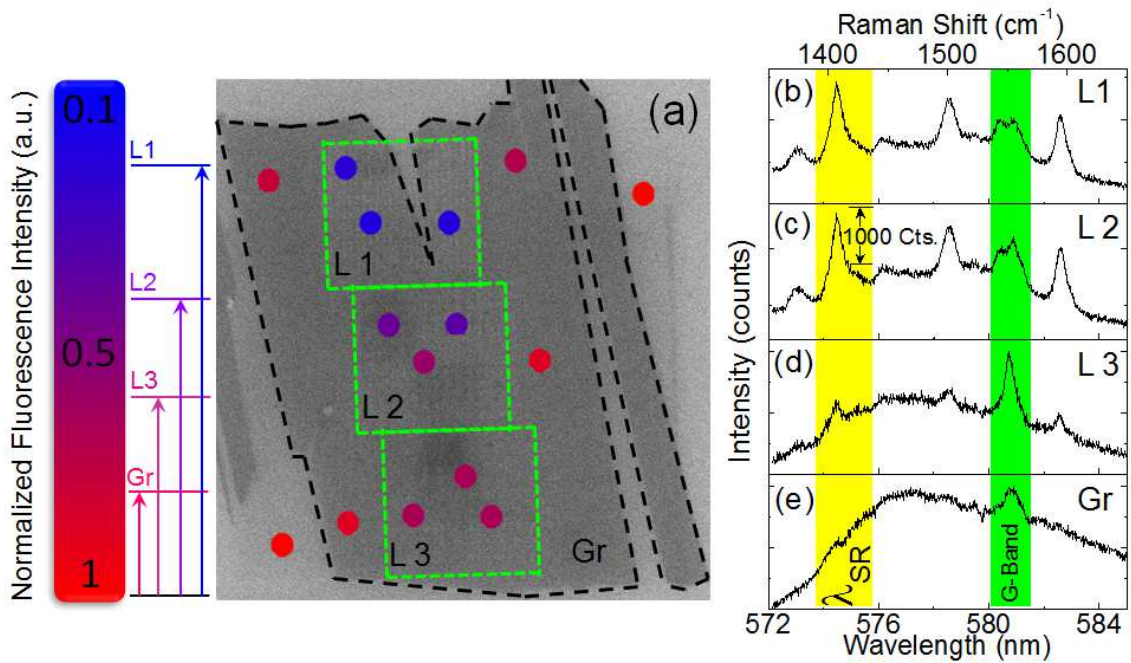}

\caption{(a) Scanning electron micrograph of the graphene flake, with nanopatterned
areas outlined by the green boxes. The filling fractions for lattice
L1, L2, and L3 are 1, 1/2, and 1/3 respectively (all dots are 100
nm in diameter), the colors correspond to the R6G fluorescence in
the sampling region normalized to the fluorescence on the bare SiO$_{2}$
wafer. SEM's of the individual lattices are available in the supporting
online materials. Several example spectra taken on (b) lattice 1,
(c) lattice 2, (d) lattice 3, and (e) unpatterned graphene, are also
shown.}

\end{figure*}

Figure 3a shows a scanning electron micrograph of a single bilayer
graphene flake with three distinct antidot superlattices L1, L2, and
L3, which was used to study the spatially resolved $\mu$-fluorescence
of the R6G dye. The relative intensity of the broad fluorescence signal
of the R6G molecule (recorded at $\lambda_{FL}=577$ nm) normalized
to the intensity of R6G fluorescence on the bare SiO$_{2}$ substrate
are identified by circles in Fig. 3a. Our results indicate that the
R6G fluorescence is moderately quenched on the unpatterned graphene
substrate as compared to the fluorescence on the bare SiO$_{2}$ wafer.
Remarkably, the fluorescence becomes even stronger quenched in the
region were the antidot superlattices are located. The amount of R6G
fluorescence quenching increases with increasing filling fraction
of the antidots as shown in Figures 3b-3e for filling fractions of
zero (graphene), 1/3 (L3), 1/2 (L2), and 1 (L1). The integrated intensity
of the R6G fluorescence signal quenches up to a factor of five for
the largest realized filling fraction, as shown in Fig 4. 

\begin{figure}
\includegraphics[scale=0.65]{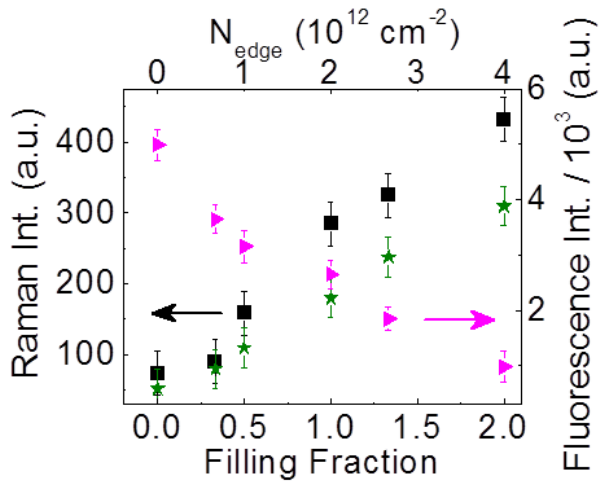}

\caption{Integrated intensity of the fluorescence signal (pink triangles) right
axis, and R6G Raman signals taken at 1390 cm$^{-1}$ (black squares)
and 1630 cm$^{-1}$ (green stars) left axis, as a function of the
antidot filling fraction; }

\end{figure}

In contrast to the quenching fluorescence signal, the intensity of
the Raman signals from both R6G and graphene were found to increase
six-fold with increasing filling fraction, i.e. increasing density
of edge states, as shown in Fig. 4. In order to rule out any possible
influence of the carboxylic bonds at the edges of the antidots and
the possible presence of oxygen groups on SiO$_{2}$, which could
have been introduced during oxygen plasma etching, a control experiment
was performed in which several antidot lattices were reduced using
1 mmol L-ascorbic acid for 24 hours. Reduction in ascorbic acid was
previously shown to effectively remove oxygen groups from graphene
\cite{Zhang_Reduction,Krauss_Zigzag_Raman}. Our results (which are
shown in the supporting online materials) indicate that no significant
oxygen contamination occurs during the 10 s etching process, and thus
cannot be used to account for the observed enhancement of the Raman
peaks. 

\begin{figure*}
\includegraphics[scale=0.72]{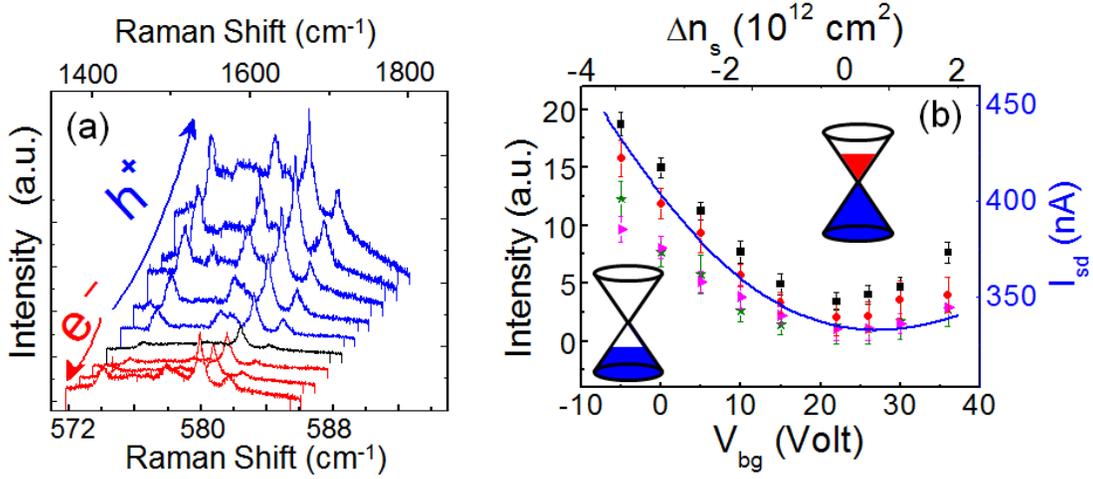}

\caption{(a) Gate tunable R6G fluorescence of an unpatterned, electrically
contacted device, similar to the one shown in Fig. 2b, and (b) intensities
of several Raman peaks (green stars taken at 1630 cm$^{-1}$and black
squares taken at 1390 cm$^{-1}$), graphene's G-band (red circles)
and R6G fluorescence (pink triangles). The blue curve shows the source-drain
current $I_{sd}$, that was used to determine the sheet carrier density
$\Delta n_{s}$ (top axis) measured in a separate transport experiment
in the same sample prior to the addition of R6G. }

\end{figure*}

Phenomenologically, the fluorescence quenching may be understood as
follows. The incident laser light creates electron-hole pairs in the
R6G dye. In the absence of the graphene substrate, the e-h pairs radiatively
recombine thereby giving rise to the fluorescence signal on the bare
SiO$_{2}$ wafer. It was previously shown that placing quantum dots
on top of graphene results in an energy transfer from the dots into
the underlying graphene layer \cite{Energy_transfer_to_Graphene},
resulting in a suppression of blinking from the quantum dots. A similar
effect is expected to occur for the R6G molecules on graphene, where
the radiative recombination of the excitons in the R6G molecule is
suppressed. In our experiments, additional quenching of the fluorescence
signal in the antidot regions was observed (as shown in Figs. 3 and
4). The additional quenching can thus be understood to arise from
the extra states at the edges $N_{edge}$, that effectively prevent
radiative recombination of the electron-hole pairs, and therefore
quench the fluorescence signal. The amount of quenching observed in
our experiments is rather remarkable since increasing the antidot
filling fraction decreases graphene's surface area and introduces
larger areas of SiO$_{2}$ into the excitation volume on which the
fluorescence is not quenched. 

The observed linear increase in carrier density with increasing filling
fraction is in accordance with the theoretical prediction of Whimmer
et al. \cite{Wimmer_antidot}, who showed that the ratio of edge states
to bulk states is given by $\frac{N_{edge}}{N_{bulk}}\approx1.07\left(1-2\alpha\right)\frac{\hbar^{2}v_{F}^{2}}{E_{0}^{2}sR}$,
where $\hbar$ is the reduced Planck's constant, $v_{F}$ is the Fermi
velocity in graphene, $\alpha$ is a parameter that characterizes
edge roughness, $E_{0}$ is the energy width of the band of edge states,
$s$ is the antidot separation, and $R$ is the antidot radius. Therefore,
decreasing $s$ or alternatively increasing $R$ gives rise to a linear
increase in $N_{edge}$.

\subsection{Gate-Tunable Fluorescence}

In order to further elucidate the mechanism for fluorescence quenching
and the nature of $N_{edge}$ we fabricated electrically contacted
and back-gated graphene flakes, which did not contain an antidot superlattice.
Varrying the backgate voltage, effectively moves the Fermi level in
the device thereby affording the possibility of in-situ electron and
hole doping of the graphene flake according to $\Delta n_{s}=C_{g}\left(V_{g}-V_{Dirac}\right)/e$,
where $C_{g}$ is the gate capacitance, $V_{g}$ is the applied gate
voltage, $V_{Dirac}$ is the location of the Dirac point, and $e$
is the electron charge \cite{Stander_Klein,Our_Klein,Graphene_Mobility,Bolotin2008}.
Modulating the Fermi level with the backgate creates a free sheet
carrier density in the underlying graphene layer. The effect of free
carriers on the R6G fluorescence and the R6G and graphene Raman is
shown in Fig. 5a, with the blue (red) traces corresponding to spectra
from hole (electron) doped regions and the black trace was taken at
the Dirac point. The intensities of several Graphene and R6G Raman
peaks are plotted in Fig. 5b together with the $I_{sd}-V_{bg}$ trace
(blue line), which illustrates that the current to the left of the
minimum (the Dirac point) is due to hole conductivity, while the current
to the right of the minimum corresponds to electron conductivity.
As can be seen, the intensities of both the Raman peaks as well as
the fluorescence signal can be either quenched or enhanced by the
applied gate bias, and directly follow the free carrier density in
the device. Comparing the values of $\Delta n_{s}$ (top axis in Fig.
5b) to $N_{edge}$ (top axis in Fig. 4) it is evident that the enhancement
of the Raman peaks achieved in antidot devices occurs at comparable
concentrations of $N_{edge}$ and sheet carrier densities $\Delta n_{s}$
in unpatterned samples, as shown in Figs. 6a and 6b. Unlike the Raman
peaks, the R6G fluorescence is strongly quenched in the nanopatterned
samples, whereas it is enhanced in the electrically gated samples.
The contrasting behavior of the fluorescence signal is strongly indicative
of the different nature of the carriers in the antidot superlattice
as compared to unpatterned graphene, and can be used to establish
a microscopic mechanism for the observed fluorescence quenching and
p-doping in the nanostructured samples.

In principle, two possible mechanisms could be responsible for fluorescence
quenching: charge transfer from R6G into the trap states that are
created by the additional edge state density or electrical field dissociation
of the radiative R6G exciton, which leads to a strong decrease in
the exciton recombination rate due to the reduced electron-hole wavefunction
overlap in an electric field. Although charge transfer into trap states
could account for the decrease of the fluorescence intensity, it cannot
explain the observed stiffening and the energetic shift of the G-band
phonon in graphene, both of which require an electric field effect
\cite{BO_Breakdown,Raman_E_Field_Tuning}. In contrast, the field
dissociation mechanism explains both phenomena, as well as the absence
of fluorescence quenching in unpatterned graphene under back-gate
sweeping. 

\begin{figure}
\includegraphics[scale=0.42]{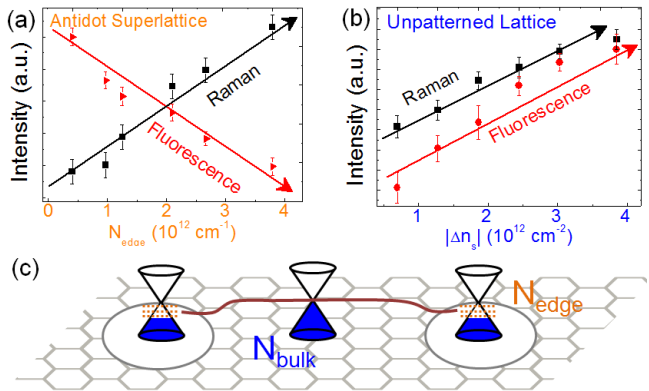}

\caption{(a) Comparison of the fluorescence quenching in the nanopatterned
samples as a function of edge state carrier density to (b) the enhancement
of fluorescence in gated samples in which free carriers are injected
into the conduction band; (c) A schematic of the band bending that
occurs as a results of pinning the Fermi level at the localized density
of states $\rho_{edge}$ at the edges of the antidots (orange dashed
lines).}

\end{figure}

Since the edge states create spatially localized carriers, which are
immobile, they would not cause the G-band stiffening. However, their
presence effectively pins the Fermi level at the edges, thereby bending
the band structure throughout the entire antidot superlattice, since
no localized states exist in graphene's basal plane and the Fermi
level must remain continuous, as shown schematically in Fig. 6c. This
band bending creates an effective potential, i.e. a built-in lateral
electrical field, that accounts for the dissociation of the R6G excitons,
resulting in the observed fluorescence quenching. In contrast, the
vertical back-gate field of the unpatterned graphene device does not
lead to band bending, while the created free carrier density can effectively
feed the carrier capture into the R6G molecules, causing the observed
fluorescence enhancement. The Raman signals are enhanced by the electrical
field mechanism providing free carriers in both cases. 

Quantitatively, the effect of the built-in electrical field may be
estimated to first order from the amount of p-doping that it introduces.
In graphene, doping is commensurate with the movement of the Fermi
level into the conduction or valence bands by the electrical field.
The band offset $\Delta E_{F}$ as a function of doping concentration
$n$ is given by $\Delta E_{F}=\hbar v_{F}k_{F}$, where $\hbar$
is the reduced Planck's constant, $v_{F}$ is the Fermi velocity,
and $k_{F}$ is the Fermi wave vector, which in graphene is given
by $k_{F}=\sqrt{\pi n}$ \cite{Hwang_carrier}. The antidot lattices
used in our experiments yielded doping concentrations on the order
of $0.5-4.0\times10^{12}$ cm$^{-2}$, which correspond to band offsets
of $\Delta E_{F}\approx90-260$ meV.

The Fermi level pinning at the localized carrier density in the antidot
superlattice is similar to Fermi level pinning of a Schottky barrier
at an graphene-metal interface that is used to separate photogenerated
carriers in optoelectronic devices based, for example, on carbon nanotubes
\cite{Avouris_CNT_Review,Contacts_1,Graphene_Photo}. In our case,
however, no metal was deposited onto graphene and the pinning occurs
at the localized edge states that are a direct consequence of the
antidot superlattice. 

In summary, we fabricated several graphene antidot superlattices using
mono, bi, and tri layer flakes, and observed effective p-type doping
which increases with larger filling fractions, as evident from their
Raman signatures. We furthermore showed that after depositing R6G
dye on these flakes, the corresponding fluorescence signal is strongly
quenched with increasing antidot filling fraction, while the Raman
signal is enhanced. These results are indicative to a microscopic
mechanism in which the Fermi level becomes pinned at the antidot periphery
giving rise to a built-in electric field, which accounts for the fluorescence
quenching and the observed p-type doping in nanopatterned graphene.
These findings make antidot lattices of great interest for carbon-based
optoelectronics and might be particularly useful for light-harvesting
applications such as photodetectors and solar cells requiring efficient
field separation of electron-hole pairs.

\bibliographystyle{achemso}
\bibliography{ANTIDOT_REVISED}

\begin{thebibliography}{42}
\expandafter\ifx\csname natexlab\endcsname\relax\def\natexlab#1{#1}\fi
\expandafter\ifx\csname bibnamefont\endcsname\relax
  \def\bibnamefont#1{#1}\fi
\expandafter\ifx\csname bibfnamefont\endcsname\relax
  \def\bibfnamefont#1{#1}\fi
\expandafter\ifx\csname citenamefont\endcsname\relax
  \def\citenamefont#1{#1}\fi
\expandafter\ifx\csname url\endcsname\relax
  \def\url#1{\texttt{#1}}\fi
\expandafter\ifx\csname urlprefix\endcsname\relax\def\urlprefix{URL }\fi
\providecommand{\bibinfo}[2]{#2}
\providecommand{\eprint}[2][]{\url{#2}}

\bibitem[{\citenamefont{Novoselov et~al.}(2004)\citenamefont{Novoselov, Geim,
  Morozov, Jiang, Zhang, Dubonos, Grigorieva, and Firsov}}]{Novoselov2004}
\bibinfo{author}{\bibfnamefont{K.~S.} \bibnamefont{Novoselov}},
  \bibinfo{author}{\bibfnamefont{A.~K.} \bibnamefont{Geim}},
  \bibinfo{author}{\bibfnamefont{S.~V.} \bibnamefont{Morozov}},
  \bibinfo{author}{\bibfnamefont{D.}~\bibnamefont{Jiang}},
  \bibinfo{author}{\bibfnamefont{Y.}~\bibnamefont{Zhang}},
  \bibinfo{author}{\bibfnamefont{S.~V.} \bibnamefont{Dubonos}},
  \bibinfo{author}{\bibfnamefont{I.~V.} \bibnamefont{Grigorieva}},
  \bibnamefont{and} \bibinfo{author}{\bibfnamefont{A.~A.}
  \bibnamefont{Firsov}}, \bibinfo{journal}{Science}
  \textbf{\bibinfo{volume}{306}}, \bibinfo{pages}{666} (\bibinfo{year}{2004}).

\bibitem[{\citenamefont{Lin et~al.}(2010)\citenamefont{Lin, Dimitrakopoulos,
  Jenkins, Farmer, Chiu, Grill, and Avouris}}]{Lin2010}
\bibinfo{author}{\bibfnamefont{Y.~M.} \bibnamefont{Lin}},
  \bibinfo{author}{\bibfnamefont{C.}~\bibnamefont{Dimitrakopoulos}},
  \bibinfo{author}{\bibfnamefont{K.~A.} \bibnamefont{Jenkins}},
  \bibinfo{author}{\bibfnamefont{D.~B.} \bibnamefont{Farmer}},
  \bibinfo{author}{\bibfnamefont{H.~Y.} \bibnamefont{Chiu}},
  \bibinfo{author}{\bibfnamefont{A.}~\bibnamefont{Grill}}, \bibnamefont{and}
  \bibinfo{author}{\bibfnamefont{P.}~\bibnamefont{Avouris}},
  \bibinfo{journal}{Science} \textbf{\bibinfo{volume}{327}},
  \bibinfo{pages}{662} (\bibinfo{year}{2010}).

\bibitem[{\citenamefont{Bolotin et~al.}(2008)\citenamefont{Bolotin, Sikes,
  Jiang, Klima, Fudenberg, Hone, Kim, and Stormer}}]{Bolotin2008}
\bibinfo{author}{\bibfnamefont{K.~I.} \bibnamefont{Bolotin}},
  \bibinfo{author}{\bibfnamefont{K.~J.} \bibnamefont{Sikes}},
  \bibinfo{author}{\bibfnamefont{Z.}~\bibnamefont{Jiang}},
  \bibinfo{author}{\bibfnamefont{M.}~\bibnamefont{Klima}},
  \bibinfo{author}{\bibfnamefont{G.}~\bibnamefont{Fudenberg}},
  \bibinfo{author}{\bibfnamefont{J.}~\bibnamefont{Hone}},
  \bibinfo{author}{\bibfnamefont{P.}~\bibnamefont{Kim}}, \bibnamefont{and}
  \bibinfo{author}{\bibfnamefont{H.~L.} \bibnamefont{Stormer}},
  \bibinfo{journal}{Solid State Communications} \textbf{\bibinfo{volume}{146}},
  \bibinfo{pages}{351} (\bibinfo{year}{2008}).

\bibitem[{\citenamefont{Balandin et~al.}(2008)\citenamefont{Balandin, Bao,
  Calizo, TeweldebrhanD, Miao, and Lau}}]{ABalandin2008}
\bibinfo{author}{\bibfnamefont{A.}~\bibnamefont{Balandin}},
  \bibinfo{author}{\bibfnamefont{W.~Z. G.~S.} \bibnamefont{Bao}},
  \bibinfo{author}{\bibfnamefont{I.}~\bibnamefont{Calizo}},
  \bibinfo{author}{\bibfnamefont{D.}~\bibnamefont{TeweldebrhanD}},
  \bibinfo{author}{\bibfnamefont{F.}~\bibnamefont{Miao}}, \bibnamefont{and}
  \bibinfo{author}{\bibfnamefont{C.~N.} \bibnamefont{Lau}},
  \bibinfo{journal}{Nano Letters} \textbf{\bibinfo{volume}{8}},
  \bibinfo{pages}{902} (\bibinfo{year}{2008}).

\bibitem[{\citenamefont{Lee et~al.}(2008)\citenamefont{Lee, Wei, Kysar, and
  Hone}}]{Hone_Durability}
\bibinfo{author}{\bibfnamefont{C.}~\bibnamefont{Lee}},
  \bibinfo{author}{\bibfnamefont{X.}~\bibnamefont{Wei}},
  \bibinfo{author}{\bibfnamefont{J.~W.} \bibnamefont{Kysar}}, \bibnamefont{and}
  \bibinfo{author}{\bibfnamefont{J.}~\bibnamefont{Hone}},
  \bibinfo{journal}{Science} \textbf{\bibinfo{volume}{321}},
  \bibinfo{pages}{385} (\bibinfo{year}{2008}).

\bibitem[{\citenamefont{Bonaccorso et~al.}(2010)\citenamefont{Bonaccorso, Sun,
  Hasan, and Ferrari}}]{Opto_Review}
\bibinfo{author}{\bibfnamefont{F.}~\bibnamefont{Bonaccorso}},
  \bibinfo{author}{\bibfnamefont{Z.}~\bibnamefont{Sun}},
  \bibinfo{author}{\bibfnamefont{T.}~\bibnamefont{Hasan}}, \bibnamefont{and}
  \bibinfo{author}{\bibfnamefont{A.~C.} \bibnamefont{Ferrari}},
  \bibinfo{journal}{Nature Photonics} \textbf{\bibinfo{volume}{4}},
  \bibinfo{pages}{611} (\bibinfo{year}{2010}).

\bibitem[{\citenamefont{Novoselov et~al.}(2007)\citenamefont{Novoselov,
  Morozov, Mohinddin, Ponomarenko, Elias, Yang, Barbolina, Blake, Booth, Jiang
  et~al.}}]{Novoselov2007}
\bibinfo{author}{\bibfnamefont{K.~S.} \bibnamefont{Novoselov}},
  \bibinfo{author}{\bibfnamefont{S.~V.} \bibnamefont{Morozov}},
  \bibinfo{author}{\bibfnamefont{T.~M.~G.} \bibnamefont{Mohinddin}},
  \bibinfo{author}{\bibfnamefont{L.~A.} \bibnamefont{Ponomarenko}},
  \bibinfo{author}{\bibfnamefont{D.~C.} \bibnamefont{Elias}},
  \bibinfo{author}{\bibfnamefont{R.}~\bibnamefont{Yang}},
  \bibinfo{author}{\bibfnamefont{I.~I.} \bibnamefont{Barbolina}},
  \bibinfo{author}{\bibfnamefont{P.}~\bibnamefont{Blake}},
  \bibinfo{author}{\bibfnamefont{T.~J.} \bibnamefont{Booth}},
  \bibinfo{author}{\bibfnamefont{D.}~\bibnamefont{Jiang}},
  \bibnamefont{et~al.}, \bibinfo{journal}{Physica Status Solidi B-basic Solid
  State Physics} \textbf{\bibinfo{volume}{244}}, \bibinfo{pages}{4106}
  (\bibinfo{year}{2007}).

\bibitem[{\citenamefont{Neto et~al.}(2009)\citenamefont{Neto, Guinea, Peres,
  Novoselov, and Geim}}]{Neto2009}
\bibinfo{author}{\bibfnamefont{A.~C.} \bibnamefont{Neto}},
  \bibinfo{author}{\bibfnamefont{F.}~\bibnamefont{Guinea}},
  \bibinfo{author}{\bibfnamefont{N.}~\bibnamefont{Peres}},
  \bibinfo{author}{\bibfnamefont{K.}~\bibnamefont{Novoselov}},
  \bibnamefont{and} \bibinfo{author}{\bibfnamefont{A.}~\bibnamefont{Geim}},
  \bibinfo{journal}{Rev. Mod. Phys} \textbf{\bibinfo{volume}{81}},
  \bibinfo{pages}{109} (\bibinfo{year}{2009}).

\bibitem[{\citenamefont{Xia et~al.}(2009)\citenamefont{Xia, Mueller, ming Lin,
  Valdes-Garcia, and Avouris}}]{Graphene_Photo}
\bibinfo{author}{\bibfnamefont{F.}~\bibnamefont{Xia}},
  \bibinfo{author}{\bibfnamefont{T.}~\bibnamefont{Mueller}},
  \bibinfo{author}{\bibfnamefont{Y.}~\bibnamefont{ming Lin}},
  \bibinfo{author}{\bibfnamefont{A.}~\bibnamefont{Valdes-Garcia}},
  \bibnamefont{and} \bibinfo{author}{\bibfnamefont{P.}~\bibnamefont{Avouris}},
  \bibinfo{journal}{Nature Nanotechnology} \textbf{\bibinfo{volume}{4}},
  \bibinfo{pages}{839} (\bibinfo{year}{2009}).

\bibitem[{\citenamefont{Nair et~al.}(2008)\citenamefont{Nair, Blake,
  Grigorenko, Novoselov, Booth, Stauber, Peres, and
  Geim}}]{Optical_Transparency}
\bibinfo{author}{\bibfnamefont{R.~R.} \bibnamefont{Nair}},
  \bibinfo{author}{\bibfnamefont{P.}~\bibnamefont{Blake}},
  \bibinfo{author}{\bibfnamefont{A.~N.} \bibnamefont{Grigorenko}},
  \bibinfo{author}{\bibfnamefont{K.~S.} \bibnamefont{Novoselov}},
  \bibinfo{author}{\bibfnamefont{T.~J.} \bibnamefont{Booth}},
  \bibinfo{author}{\bibfnamefont{T.}~\bibnamefont{Stauber}},
  \bibinfo{author}{\bibfnamefont{N.~M.~R.} \bibnamefont{Peres}},
  \bibnamefont{and} \bibinfo{author}{\bibfnamefont{A.~K.} \bibnamefont{Geim}},
  \bibinfo{journal}{Science} \textbf{\bibinfo{volume}{320}},
  \bibinfo{pages}{5881} (\bibinfo{year}{2008}).

\bibitem[{\citenamefont{Gierz et~al.}(2007)\citenamefont{Gierz, Riedl, Starke,
  Ast, and Kern}}]{Atomic_Doping}
\bibinfo{author}{\bibfnamefont{I.}~\bibnamefont{Gierz}},
  \bibinfo{author}{\bibfnamefont{C.}~\bibnamefont{Riedl}},
  \bibinfo{author}{\bibfnamefont{U.}~\bibnamefont{Starke}},
  \bibinfo{author}{\bibfnamefont{C.~R.} \bibnamefont{Ast}}, \bibnamefont{and}
  \bibinfo{author}{\bibfnamefont{K.}~\bibnamefont{Kern}},
  \bibinfo{journal}{Nano Letters} \textbf{\bibinfo{volume}{8}},
  \bibinfo{pages}{4603} (\bibinfo{year}{2007}).

\bibitem[{\citenamefont{Wang et~al.}(2010)\citenamefont{Wang, Shao, Matson, Li,
  and Lin}}]{N_Doping_2}
\bibinfo{author}{\bibfnamefont{Y.}~\bibnamefont{Wang}},
  \bibinfo{author}{\bibfnamefont{Y.}~\bibnamefont{Shao}},
  \bibinfo{author}{\bibfnamefont{D.~W.} \bibnamefont{Matson}},
  \bibinfo{author}{\bibfnamefont{J.}~\bibnamefont{Li}}, \bibnamefont{and}
  \bibinfo{author}{\bibfnamefont{Y.}~\bibnamefont{Lin}}, \bibinfo{journal}{ACS
  Nano} \textbf{\bibinfo{volume}{4}}, \bibinfo{pages}{1790}
  (\bibinfo{year}{2010}).

\bibitem[{\citenamefont{Wehling et~al.}(2008)\citenamefont{Wehling, Novoselov,
  Morozov, Vdovin, Katsnelson, Geim, and Lichtenstein}}]{Molecular_Doping}
\bibinfo{author}{\bibfnamefont{T.~O.} \bibnamefont{Wehling}},
  \bibinfo{author}{\bibfnamefont{K.~S.} \bibnamefont{Novoselov}},
  \bibinfo{author}{\bibfnamefont{S.~V.} \bibnamefont{Morozov}},
  \bibinfo{author}{\bibfnamefont{E.~E.} \bibnamefont{Vdovin}},
  \bibinfo{author}{\bibfnamefont{M.~I.} \bibnamefont{Katsnelson}},
  \bibinfo{author}{\bibfnamefont{A.~K.} \bibnamefont{Geim}}, \bibnamefont{and}
  \bibinfo{author}{\bibfnamefont{A.~I.} \bibnamefont{Lichtenstein}},
  \bibinfo{journal}{Nano Letters} \textbf{\bibinfo{volume}{8}},
  \bibinfo{pages}{173} (\bibinfo{year}{2008}).

\bibitem[{\citenamefont{Dong et~al.}(2009)\citenamefont{Dong, Fu, Fang, Shi,
  Chen, and Li}}]{Molecular_Doping_2}
\bibinfo{author}{\bibfnamefont{X.}~\bibnamefont{Dong}},
  \bibinfo{author}{\bibfnamefont{D.}~\bibnamefont{Fu}},
  \bibinfo{author}{\bibfnamefont{W.}~\bibnamefont{Fang}},
  \bibinfo{author}{\bibfnamefont{Y.}~\bibnamefont{Shi}},
  \bibinfo{author}{\bibfnamefont{P.}~\bibnamefont{Chen}}, \bibnamefont{and}
  \bibinfo{author}{\bibfnamefont{L.-J.} \bibnamefont{Li}},
  \bibinfo{journal}{Small} \textbf{\bibinfo{volume}{5}}, \bibinfo{pages}{1422}
  (\bibinfo{year}{2009}).

\bibitem[{\citenamefont{Guo et~al.}(2010)\citenamefont{Guo, Liu, Chen, Zhu,
  Fang, and Gong}}]{Nitrogen_Doping}
\bibinfo{author}{\bibfnamefont{B.}~\bibnamefont{Guo}},
  \bibinfo{author}{\bibfnamefont{Q.}~\bibnamefont{Liu}},
  \bibinfo{author}{\bibfnamefont{E.}~\bibnamefont{Chen}},
  \bibinfo{author}{\bibfnamefont{H.}~\bibnamefont{Zhu}},
  \bibinfo{author}{\bibfnamefont{L.}~\bibnamefont{Fang}}, \bibnamefont{and}
  \bibinfo{author}{\bibfnamefont{J.~R.} \bibnamefont{Gong}},
  \bibinfo{journal}{Nano Letters} \textbf{\bibinfo{volume}{In Press}}
  (\bibinfo{year}{2010}).

\bibitem[{\citenamefont{Mueller et~al.}(2009)\citenamefont{Mueller, Xia,
  Freitag, Tsang, and Avouris}}]{Contacts_1}
\bibinfo{author}{\bibfnamefont{T.}~\bibnamefont{Mueller}},
  \bibinfo{author}{\bibfnamefont{F.}~\bibnamefont{Xia}},
  \bibinfo{author}{\bibfnamefont{M.}~\bibnamefont{Freitag}},
  \bibinfo{author}{\bibfnamefont{J.}~\bibnamefont{Tsang}}, \bibnamefont{and}
  \bibinfo{author}{\bibfnamefont{P.}~\bibnamefont{Avouris}},
  \bibinfo{journal}{Phys. Rev. B} \textbf{\bibinfo{volume}{79}},
  \bibinfo{pages}{245430} (\bibinfo{year}{2009}).

\bibitem[{\citenamefont{Mueller et~al.}(2010)\citenamefont{Mueller, Xia, and
  Avouris}}]{Graphene_Photo_2}
\bibinfo{author}{\bibfnamefont{T.}~\bibnamefont{Mueller}},
  \bibinfo{author}{\bibfnamefont{F.}~\bibnamefont{Xia}}, \bibnamefont{and}
  \bibinfo{author}{\bibfnamefont{P.}~\bibnamefont{Avouris}},
  \bibinfo{journal}{Nature Photonics} \textbf{\bibinfo{volume}{4}},
  \bibinfo{pages}{297} (\bibinfo{year}{2010}).

\bibitem[{\citenamefont{Avouris}(2006)}]{Avouris_CNT_Review}
\bibinfo{author}{\bibfnamefont{P.}~\bibnamefont{Avouris}},
  \bibinfo{journal}{Materials Today} \textbf{\bibinfo{volume}{9}},
  \bibinfo{pages}{46} (\bibinfo{year}{2006}).

\bibitem[{\citenamefont{Hong et~al.}(2010)\citenamefont{Hong, Tabakman,
  Welsher, Wang, Wang, and Dai}}]{Hong_CNT}
\bibinfo{author}{\bibfnamefont{G.}~\bibnamefont{Hong}},
  \bibinfo{author}{\bibfnamefont{S.~M.} \bibnamefont{Tabakman}},
  \bibinfo{author}{\bibfnamefont{K.}~\bibnamefont{Welsher}},
  \bibinfo{author}{\bibfnamefont{H.}~\bibnamefont{Wang}},
  \bibinfo{author}{\bibfnamefont{X.}~\bibnamefont{Wang}}, \bibnamefont{and}
  \bibinfo{author}{\bibfnamefont{H.}~\bibnamefont{Dai}}, \bibinfo{journal}{J.
  Am. Chem. Soc.} \textbf{\bibinfo{volume}{132}}, \bibinfo{pages}{15920}
  (\bibinfo{year}{2010}).

\bibitem[{\citenamefont{Chen et~al.}(2010{\natexlab{a}})\citenamefont{Chen,
  Berciaud, Nuckolls, Heinz, and Brus}}]{Energy_Transfer}
\bibinfo{author}{\bibfnamefont{Z.}~\bibnamefont{Chen}},
  \bibinfo{author}{\bibfnamefont{S.}~\bibnamefont{Berciaud}},
  \bibinfo{author}{\bibfnamefont{C.}~\bibnamefont{Nuckolls}},
  \bibinfo{author}{\bibfnamefont{T.~F.} \bibnamefont{Heinz}}, \bibnamefont{and}
  \bibinfo{author}{\bibfnamefont{L.~E.} \bibnamefont{Brus}},
  \bibinfo{journal}{ACS Nano} \textbf{\bibinfo{volume}{4}},
  \bibinfo{pages}{2964} (\bibinfo{year}{2010}{\natexlab{a}}).

\bibitem[{\citenamefont{Wimmer et~al.}(2010)\citenamefont{Wimmer, Akhmerov, and
  Guinea}}]{Wimmer_antidot}
\bibinfo{author}{\bibfnamefont{M.}~\bibnamefont{Wimmer}},
  \bibinfo{author}{\bibfnamefont{A.~R.} \bibnamefont{Akhmerov}},
  \bibnamefont{and} \bibinfo{author}{\bibfnamefont{F.}~\bibnamefont{Guinea}},
  \bibinfo{journal}{Phys. Rev. B} \textbf{\bibinfo{volume}{82}},
  \bibinfo{pages}{045409} (\bibinfo{year}{2010}).

\bibitem[{\citenamefont{Bai et~al.}(2010)\citenamefont{Bai, Zhong, Jiang,
  Huang, and Duan}}]{Bai_nanomesh}
\bibinfo{author}{\bibfnamefont{J.}~\bibnamefont{Bai}},
  \bibinfo{author}{\bibfnamefont{X.}~\bibnamefont{Zhong}},
  \bibinfo{author}{\bibfnamefont{S.}~\bibnamefont{Jiang}},
  \bibinfo{author}{\bibfnamefont{Y.}~\bibnamefont{Huang}}, \bibnamefont{and}
  \bibinfo{author}{\bibfnamefont{X.}~\bibnamefont{Duan}},
  \bibinfo{journal}{Nature Nanotechnol.} \textbf{\bibinfo{volume}{5}},
  \bibinfo{pages}{190} (\bibinfo{year}{2010}).

\bibitem[{\citenamefont{Liang et~al.}(2010)\citenamefont{Liang, Jung, Wu,
  Ismach, Olynick, Cabrini, and Bokor}}]{Liang_BandGap}
\bibinfo{author}{\bibfnamefont{X.}~\bibnamefont{Liang}},
  \bibinfo{author}{\bibfnamefont{Y.-S.} \bibnamefont{Jung}},
  \bibinfo{author}{\bibfnamefont{S.}~\bibnamefont{Wu}},
  \bibinfo{author}{\bibfnamefont{A.}~\bibnamefont{Ismach}},
  \bibinfo{author}{\bibfnamefont{D.~L.} \bibnamefont{Olynick}},
  \bibinfo{author}{\bibfnamefont{S.}~\bibnamefont{Cabrini}}, \bibnamefont{and}
  \bibinfo{author}{\bibfnamefont{J.}~\bibnamefont{Bokor}},
  \bibinfo{journal}{Nano Lett.} \textbf{\bibinfo{volume}{10}},
  \bibinfo{pages}{2454} (\bibinfo{year}{2010}).

\bibitem[{\citenamefont{Kim et~al.}(2010)\citenamefont{Kim, Safron, Han,
  Arnold, and Gopalan}}]{Kim_Nanoperf}
\bibinfo{author}{\bibfnamefont{M.}~\bibnamefont{Kim}},
  \bibinfo{author}{\bibfnamefont{N.~S.} \bibnamefont{Safron}},
  \bibinfo{author}{\bibfnamefont{E.}~\bibnamefont{Han}},
  \bibinfo{author}{\bibfnamefont{M.~S.} \bibnamefont{Arnold}},
  \bibnamefont{and} \bibinfo{author}{\bibfnamefont{P.}~\bibnamefont{Gopalan}},
  \bibinfo{journal}{Nano Lett.} \textbf{\bibinfo{volume}{10}},
  \bibinfo{pages}{1125} (\bibinfo{year}{2010}).

\bibitem[{\citenamefont{Sinitskii and Tour}(2010)}]{Sinitskii_Patterning}
\bibinfo{author}{\bibfnamefont{A.}~\bibnamefont{Sinitskii}} \bibnamefont{and}
  \bibinfo{author}{\bibfnamefont{J.~M.} \bibnamefont{Tour}},
  \bibinfo{journal}{J. Am. Chem. Soc.} \textbf{\bibinfo{volume}{132}},
  \bibinfo{pages}{14730} (\bibinfo{year}{2010}).

\bibitem[{\citenamefont{Stojanovic et~al.}(2010)\citenamefont{Stojanovic,
  Vukmirovic, and Bruder}}]{antidot_polaron}
\bibinfo{author}{\bibfnamefont{V.~M.} \bibnamefont{Stojanovic}},
  \bibinfo{author}{\bibfnamefont{N.}~\bibnamefont{Vukmirovic}},
  \bibnamefont{and} \bibinfo{author}{\bibfnamefont{C.}~\bibnamefont{Bruder}},
  \bibinfo{journal}{Phys. Rev. B} \textbf{\bibinfo{volume}{82}},
  \bibinfo{pages}{165410} (\bibinfo{year}{2010}).

\bibitem[{\citenamefont{Petersen and Pedersen}(2009)}]{Antidot_quasiparticles}
\bibinfo{author}{\bibfnamefont{R.}~\bibnamefont{Petersen}} \bibnamefont{and}
  \bibinfo{author}{\bibfnamefont{T.~G.} \bibnamefont{Pedersen}},
  \bibinfo{journal}{Phys. Rev. B} \textbf{\bibinfo{volume}{80}},
  \bibinfo{pages}{113404} (\bibinfo{year}{2009}).

\bibitem[{\citenamefont{Heydrich et~al.}(2010)\citenamefont{Heydrich, Hirmer,
  Preis, Korn, Eroms, Weiss, and Schuller}}]{Heydrich_antidot_apl}
\bibinfo{author}{\bibfnamefont{S.}~\bibnamefont{Heydrich}},
  \bibinfo{author}{\bibfnamefont{M.}~\bibnamefont{Hirmer}},
  \bibinfo{author}{\bibfnamefont{C.}~\bibnamefont{Preis}},
  \bibinfo{author}{\bibfnamefont{T.}~\bibnamefont{Korn}},
  \bibinfo{author}{\bibfnamefont{J.}~\bibnamefont{Eroms}},
  \bibinfo{author}{\bibfnamefont{D.}~\bibnamefont{Weiss}}, \bibnamefont{and}
  \bibinfo{author}{\bibfnamefont{C.}~\bibnamefont{Schuller}},
  \bibinfo{journal}{Appl. Phys. Lett.} \textbf{\bibinfo{volume}{97}},
  \bibinfo{pages}{043113} (\bibinfo{year}{2010}).

\bibitem[{\citenamefont{Begliarbekov
  et~al.}(2010{\natexlab{a}})\citenamefont{Begliarbekov, Sul, Kalliakos, Yang,
  and Strauf}}]{Begliarbekov2010}
\bibinfo{author}{\bibfnamefont{M.}~\bibnamefont{Begliarbekov}},
  \bibinfo{author}{\bibfnamefont{O.}~\bibnamefont{Sul}},
  \bibinfo{author}{\bibfnamefont{S.}~\bibnamefont{Kalliakos}},
  \bibinfo{author}{\bibfnamefont{E.-H.} \bibnamefont{Yang}}, \bibnamefont{and}
  \bibinfo{author}{\bibfnamefont{S.}~\bibnamefont{Strauf}},
  \bibinfo{journal}{Appl. Phys. Lett.} \textbf{\bibinfo{volume}{97}},
  \bibinfo{pages}{031908} (\bibinfo{year}{2010}{\natexlab{a}}).

\bibitem[{\citenamefont{Ferrari et~al.}(2006)\citenamefont{Ferrari, Meyer,
  Scardaci, Casiraghi, Lazzeri, Mauri, Piscanec, Jiang, S., Roth
  et~al.}}]{Graphene_Raman}
\bibinfo{author}{\bibfnamefont{A.~C.} \bibnamefont{Ferrari}},
  \bibinfo{author}{\bibfnamefont{J.~C.} \bibnamefont{Meyer}},
  \bibinfo{author}{\bibfnamefont{V.}~\bibnamefont{Scardaci}},
  \bibinfo{author}{\bibfnamefont{C.}~\bibnamefont{Casiraghi}},
  \bibinfo{author}{\bibfnamefont{M.}~\bibnamefont{Lazzeri}},
  \bibinfo{author}{\bibfnamefont{F.}~\bibnamefont{Mauri}},
  \bibinfo{author}{\bibfnamefont{S.}~\bibnamefont{Piscanec}},
  \bibinfo{author}{\bibfnamefont{D.}~\bibnamefont{Jiang}},
  \bibinfo{author}{\bibfnamefont{K.~S.~N.} \bibnamefont{S.}},
  \bibinfo{author}{\bibnamefont{Roth}}, \bibnamefont{et~al.},
  \bibinfo{journal}{Phys. Rev. Lett} \textbf{\bibinfo{volume}{97}},
  \bibinfo{pages}{187401} (\bibinfo{year}{2006}).

\bibitem[{\citenamefont{Yan et~al.}(2007)\citenamefont{Yan, Zhang, Kim, and
  Pinczuk}}]{Raman_E_Field_Tuning}
\bibinfo{author}{\bibfnamefont{J.}~\bibnamefont{Yan}},
  \bibinfo{author}{\bibfnamefont{Y.}~\bibnamefont{Zhang}},
  \bibinfo{author}{\bibfnamefont{P.}~\bibnamefont{Kim}}, \bibnamefont{and}
  \bibinfo{author}{\bibfnamefont{A.}~\bibnamefont{Pinczuk}},
  \bibinfo{journal}{Phys. Rev. Lett} \textbf{\bibinfo{volume}{98}},
  \bibinfo{pages}{166802} (\bibinfo{year}{2007}).

\bibitem[{\citenamefont{Pisana et~al.}(2007)\citenamefont{Pisana, Lazzeri,
  Casiraghi, Novoselov, Geim, Ferrari, and Mauri}}]{BO_Breakdown}
\bibinfo{author}{\bibfnamefont{S.}~\bibnamefont{Pisana}},
  \bibinfo{author}{\bibfnamefont{M.}~\bibnamefont{Lazzeri}},
  \bibinfo{author}{\bibfnamefont{C.}~\bibnamefont{Casiraghi}},
  \bibinfo{author}{\bibfnamefont{K.~S.} \bibnamefont{Novoselov}},
  \bibinfo{author}{\bibfnamefont{A.~K.} \bibnamefont{Geim}},
  \bibinfo{author}{\bibfnamefont{A.~C.} \bibnamefont{Ferrari}},
  \bibnamefont{and} \bibinfo{author}{\bibfnamefont{F.}~\bibnamefont{Mauri}},
  \bibinfo{journal}{Nature Materials} \textbf{\bibinfo{volume}{6}},
  \bibinfo{pages}{198 } (\bibinfo{year}{2007}).

\bibitem[{\citenamefont{Ferrari}(2007)}]{Ferrari2007}
\bibinfo{author}{\bibfnamefont{A.~C.} \bibnamefont{Ferrari}},
  \bibinfo{journal}{Solid State Communications} \textbf{\bibinfo{volume}{143}},
  \bibinfo{pages}{47} (\bibinfo{year}{2007}).

\bibitem[{\citenamefont{Das et~al.}(2008)\citenamefont{Das, Pisana,
  Chakraborty, Piscanec, Saha, Waghmare, Novoselov, Krishnamurthy, Geim,
  Ferrari et~al.}}]{Gated_Raman}
\bibinfo{author}{\bibfnamefont{A.}~\bibnamefont{Das}},
  \bibinfo{author}{\bibfnamefont{S.}~\bibnamefont{Pisana}},
  \bibinfo{author}{\bibfnamefont{B.}~\bibnamefont{Chakraborty}},
  \bibinfo{author}{\bibfnamefont{S.}~\bibnamefont{Piscanec}},
  \bibinfo{author}{\bibfnamefont{S.~K.} \bibnamefont{Saha}},
  \bibinfo{author}{\bibfnamefont{U.~V.} \bibnamefont{Waghmare}},
  \bibinfo{author}{\bibfnamefont{K.~S.} \bibnamefont{Novoselov}},
  \bibinfo{author}{\bibfnamefont{H.~R.} \bibnamefont{Krishnamurthy}},
  \bibinfo{author}{\bibfnamefont{A.~K.} \bibnamefont{Geim}},
  \bibinfo{author}{\bibfnamefont{A.~C.} \bibnamefont{Ferrari}},
  \bibnamefont{et~al.}, \bibinfo{journal}{Nature Nanotechnology}
  \textbf{\bibinfo{volume}{3}}, \bibinfo{pages}{210 } (\bibinfo{year}{2008}).

\bibitem[{\citenamefont{Stampfer et~al.}(2007)\citenamefont{Stampfer, Molitor,
  Graf, Ensslin, Jungen, Hierold, and Wirtz}}]{Raman_Doping_Stampfer}
\bibinfo{author}{\bibfnamefont{C.}~\bibnamefont{Stampfer}},
  \bibinfo{author}{\bibfnamefont{F.}~\bibnamefont{Molitor}},
  \bibinfo{author}{\bibfnamefont{D.}~\bibnamefont{Graf}},
  \bibinfo{author}{\bibfnamefont{K.}~\bibnamefont{Ensslin}},
  \bibinfo{author}{\bibfnamefont{A.}~\bibnamefont{Jungen}},
  \bibinfo{author}{\bibfnamefont{C.}~\bibnamefont{Hierold}}, \bibnamefont{and}
  \bibinfo{author}{\bibfnamefont{L.}~\bibnamefont{Wirtz}},
  \bibinfo{journal}{Appl. Phys. Lett.} \textbf{\bibinfo{volume}{91}},
  \bibinfo{pages}{241907} (\bibinfo{year}{2007}).

\bibitem[{\citenamefont{Zhang et~al.}(2010)\citenamefont{Zhang, Yang, Shen,
  Cheng, Zhang, and Guo}}]{Zhang_Reduction}
\bibinfo{author}{\bibfnamefont{J.}~\bibnamefont{Zhang}},
  \bibinfo{author}{\bibfnamefont{H.}~\bibnamefont{Yang}},
  \bibinfo{author}{\bibfnamefont{G.}~\bibnamefont{Shen}},
  \bibinfo{author}{\bibfnamefont{P.}~\bibnamefont{Cheng}},
  \bibinfo{author}{\bibfnamefont{J.}~\bibnamefont{Zhang}}, \bibnamefont{and}
  \bibinfo{author}{\bibfnamefont{S.}~\bibnamefont{Guo}},
  \bibinfo{journal}{Chem. Commun.} \textbf{\bibinfo{volume}{46}},
  \bibinfo{pages}{1112} (\bibinfo{year}{2010}).

\bibitem[{\citenamefont{Krauss et~al.}(2010)\citenamefont{Krauss, Nemes-Incze,
  Skakalova, Biro, von Klitzing, and Smet}}]{Krauss_Zigzag_Raman}
\bibinfo{author}{\bibfnamefont{B.}~\bibnamefont{Krauss}},
  \bibinfo{author}{\bibfnamefont{P.}~\bibnamefont{Nemes-Incze}},
  \bibinfo{author}{\bibfnamefont{V.}~\bibnamefont{Skakalova}},
  \bibinfo{author}{\bibfnamefont{L.~P.} \bibnamefont{Biro}},
  \bibinfo{author}{\bibfnamefont{K.}~\bibnamefont{von Klitzing}},
  \bibnamefont{and} \bibinfo{author}{\bibfnamefont{J.~H.} \bibnamefont{Smet}},
  \bibinfo{journal}{Nano Lett.} \textbf{\bibinfo{volume}{10}},
  \bibinfo{pages}{4544} (\bibinfo{year}{2010}).

\bibitem[{\citenamefont{Chen et~al.}(2010{\natexlab{b}})\citenamefont{Chen,
  Berciaud, Nuckolls, Heinz, and Brus}}]{Energy_transfer_to_Graphene}
\bibinfo{author}{\bibfnamefont{Z.}~\bibnamefont{Chen}},
  \bibinfo{author}{\bibfnamefont{S.}~\bibnamefont{Berciaud}},
  \bibinfo{author}{\bibfnamefont{C.}~\bibnamefont{Nuckolls}},
  \bibinfo{author}{\bibfnamefont{T.~F.} \bibnamefont{Heinz}}, \bibnamefont{and}
  \bibinfo{author}{\bibfnamefont{L.~E.} \bibnamefont{Brus}},
  \bibinfo{journal}{ACS Nano} \textbf{\bibinfo{volume}{4}},
  \bibinfo{pages}{2964} (\bibinfo{year}{2010}{\natexlab{b}}).

\bibitem[{\citenamefont{Stander et~al.}(2009)\citenamefont{Stander, Huard, and
  Goldhaber-Gordon}}]{Stander_Klein}
\bibinfo{author}{\bibfnamefont{N.}~\bibnamefont{Stander}},
  \bibinfo{author}{\bibfnamefont{B.}~\bibnamefont{Huard}}, \bibnamefont{and}
  \bibinfo{author}{\bibfnamefont{D.}~\bibnamefont{Goldhaber-Gordon}},
  \bibinfo{journal}{Phys. Rev. Lett.} \textbf{\bibinfo{volume}{102}},
  \bibinfo{pages}{026807} (\bibinfo{year}{2009}).

\bibitem[{\citenamefont{Begliarbekov
  et~al.}(2010{\natexlab{b}})\citenamefont{Begliarbekov, Sul, Ai, Yang, and
  Strauf}}]{Our_Klein}
\bibinfo{author}{\bibfnamefont{M.}~\bibnamefont{Begliarbekov}},
  \bibinfo{author}{\bibfnamefont{O.}~\bibnamefont{Sul}},
  \bibinfo{author}{\bibfnamefont{N.}~\bibnamefont{Ai}},
  \bibinfo{author}{\bibfnamefont{E.-H.} \bibnamefont{Yang}}, \bibnamefont{and}
  \bibinfo{author}{\bibfnamefont{S.}~\bibnamefont{Strauf}},
  \bibinfo{journal}{Appl. Phys. Lett.} \textbf{\bibinfo{volume}{97}},
  \bibinfo{pages}{122106} (\bibinfo{year}{2010}{\natexlab{b}}).

\bibitem[{\citenamefont{Tan et~al.}(2007)\citenamefont{Tan, Zhang, Bolotin,
  Zhao, Adam, Hwang, Sarma, Stormer, and Kim}}]{Graphene_Mobility}
\bibinfo{author}{\bibfnamefont{Y.-W.} \bibnamefont{Tan}},
  \bibinfo{author}{\bibfnamefont{Y.}~\bibnamefont{Zhang}},
  \bibinfo{author}{\bibfnamefont{K.}~\bibnamefont{Bolotin}},
  \bibinfo{author}{\bibfnamefont{Y.}~\bibnamefont{Zhao}},
  \bibinfo{author}{\bibfnamefont{S.}~\bibnamefont{Adam}},
  \bibinfo{author}{\bibfnamefont{E.~H.} \bibnamefont{Hwang}},
  \bibinfo{author}{\bibfnamefont{S.~D.} \bibnamefont{Sarma}},
  \bibinfo{author}{\bibfnamefont{H.~L.} \bibnamefont{Stormer}},
  \bibnamefont{and} \bibinfo{author}{\bibfnamefont{P.}~\bibnamefont{Kim}},
  \bibinfo{journal}{Phys. Rev. Lett.} \textbf{\bibinfo{volume}{99}},
  \bibinfo{pages}{246803} (\bibinfo{year}{2007}).

\bibitem[{\citenamefont{Hwang et~al.}(2007)\citenamefont{Hwang, Adam, and
  Sarma}}]{Hwang_carrier}
\bibinfo{author}{\bibfnamefont{E.}~\bibnamefont{Hwang}},
  \bibinfo{author}{\bibfnamefont{S.}~\bibnamefont{Adam}}, \bibnamefont{and}
  \bibinfo{author}{\bibfnamefont{S.~D.} \bibnamefont{Sarma}},
  \bibinfo{journal}{Phys. Rev. Lett.} \textbf{\bibinfo{volume}{98}},
  \bibinfo{pages}{186806} (\bibinfo{year}{2007}).

\end{thebibliography}

\end{document}